\documentclass[aps,preprint,12pt]{revtex4}
\usepackage{graphicx}
\usepackage{longtable}
\begin{document}
\title{Resonance phenomena in ultracold dipole-dipole scattering}

\author{V. Roudnev\footnote{roudnev@pa.uky.edu} and M. Cavagnero}

\address{Department of Physics and Astronomy, University of Kentucky, 40506,KY USA}

\begin{abstract}
Elastic scattering resonances occurring in ultracold collisions of either bosonic or fermionic polar molecules are investigated. The Born-Oppenheimer adiabatic representation
of the two-body dynamics provides both a qualitative classification scheme and a quantitative WKB quantization condition that predicts several 
sequences of resonant states. 
It is found that the near-threshold energy dependence of ultracold collision cross sections varies significantly with the particle exchange symmetry, with bosonic systems
showing much smoother energy variations than their fermionic counterparts.   
Resonant variations of the angular distributions in ultracold collisions are also described. 
\end{abstract}
\maketitle
\noindent \textit{Keywords}: {dipoles, low-energy scattering, resonances} 
\maketitle

\section{Introduction}
While numerous advances have been made in the study of dilute atomic gases at ultracold temperatures, very little is yet known about ultracold molecular gases, particularly for strongly polar molecules. 
The first experiments with ensembles of trapped polar molecules are only recently reported~\cite{JunYe}.
Molecular gases are more difficult to cool than their atomic counterparts, and the temperatures at which effects of quantum degeneracy are apparent are typically much lower.
Improved cooling and trapping techniques for cold molecules have emerged, including the development  of decelerators~\cite{decelerators} and the formation of molecules from ultracold atoms via photoassociation~\cite{photoassociation}.  
But most trapped molecular ensembles are still relatively ``hot'' and rather far from the quantum degenerate regime. 

Many theoretical studies of molecular collisions at cold and ultracold temperatures are now emerging in order to spur and assist these ongoing experimental efforts~\cite{Review1,theory,theory1,theory2,theory3}. 
One of the most striking outcomes of recent studies is the identification of large elastic scattering resonances in zero-energy collisions of virtually all molecules aligned in an external electric field~\cite{TickBohn1}. 
These resonances can be ``tuned'' with the electric field strength, resulting in a novel kind of  zero-energy spectroscopy~\cite{TickBohn2}, and they have also been predicted in calculations of trapped ensembles~\cite{Doerte}. 
In an earlier work, we demonstrated the universality of these resonances for bosonic collision partners~\cite{ourPRL}. 
In this paper, we provide a number of details of the resonance classification, and extend the work by comparing and contrasting
with similar resonances in collisions between fermionic polar molecules.  

A few general observations highlight the importance that resonance formation and decay will play in the
dynamics of ultracold molecular gases: 1) near-threshold collisions of polar molecules are ``resonance rich'',
with resonances of shape, Feshbach/Fano and mixed characteristics forming in the long-range dipole-dipole
field, and with detailed characteristics sensitive to shorter ranged interactions; 2) the resonances occur
systematically and can readily be predicted by simple semi-classical quantization formulas; 3) the character
and frequency of resonances differ distinctly between bose and fermi collision partners; 4) the angular 
distribution of scattered products in near-zero energy collisions is strongly correlated with resonance 
formation and decay; and 5) the resonances are universal and, with minor field variations, can be realized
in virtually all strongly polar molecules. This ubiquity of threshold resonance phenomena  
implies that complex few and many-body processes in ultracold polar gases are likely to emerge as experiments progress to the
quantum-degenerate regime.  

\section{Long-range interaction of polar molecules}
There are many molecules (such as alkali-halide salts) which, in their lowest energy states, exhibit strong polar characteristics. 
However, in the absence of external fields, as a consequence of parity conservation, even these strongly "polar" molecules have zero dipole moment. 
Molecular interactions may polarize such collision partners, even in the absence of any external fields, though this generally produces a relatively benign and isotropic ($1/r^6$) van der Waals attraction at large intermolecular separations. 
The stronger ($1/r^3$) anisotropic interactions of interest here require the imposition of an external electric field, $\vec{\cal E}$, coupling opposite parity states, and producing a generally field-dependent dipole moment  
\begin{eqnarray} 
   \vec \mu({\cal E}) = -(\frac{dE}{d{\cal E}})\hat {\cal E}
\end{eqnarray}
for all molecules. 

The magnitude and sign of the "induced" dipole moment, $\mu$, therefore depends on specific details of the $E({\cal E})$ Stark shift of the state of interest for any particular molecule~\cite{BohnStark}. 
In general, for zero field, two molecular states of opposite parity are separated by an energy gap, so that $E({\cal E})$ varies quadratically for
small fields. 
At larger fields, Stark maps tend to a linear regime, yielding a maximal induced dipole moment $\vec{\mu}\rightarrow\vec{d}=qs\hat{d}$,
where $q$ is some effective charge, and $s$ is a measure of the charge separation along the field axis, and typically is a few Bohr in size. 
The magnitude of the electric field required to maximize the dipole moment $\mu$ depends principally on the size of the zero-field energy gap; molecules with ground $\Sigma$ states tend to have relatively large gaps and often require experimentally unrealistic field strengths, while those in $\Pi$ or $\Delta$  states can have quite small gaps (owing to the phenomenon of $\Lambda$-doubling), producing maximal dipole moments with modest fields.

For ground vibrational states of $\Sigma$ molecules $\mu({\cal E})$ can be estimated as
\begin{equation} 
  \vec \mu = -({dE\over d{\cal E}})\hat z = - d^2\frac{\cal E}{3B}\hat{z} \ \ \ ,
\end{equation}
where $B$ is the rotational constant of the molecule and $\hat{z}$ is the field direction.  
In Table~\ref{Tab:SigmaMol} we provide typical values of the intrinsic dipole moment $d$ and the 
rotational constant $B$ for a variety of $\Sigma$-state molecules~\cite{NIST}. 
\begin{longtable}{ccccc}
\caption{\label{Tab:SigmaMol} Important dipolar and rotational constants of $\Sigma$ molecules. $2.928(-05)$ stands for $2.928\times 10^{-05}$. The dipole length, 
$D$ is calculated in the assumption that the molecule is completely polarized.}\\
Molecule			&   $B_0$,a.u.	&   $d$, a.u.	&   $M/m_e$	&   $D$	\\
HI						&  2.928(-05)	&  0.1762	&  116584.923	&  3618	\\
H$^{35}$Cl		&  4.757(-05)	&  0.4362	&  32790.736	&  6239	\\
H$^{37}$Cl		&  4.750(-05)	&  0.4359	&  34610.935	&  6578	\\
H$^{79}$Br		&  3.805(-05)	&  0.3254	&  72848.241	&  7715	\\
H$^{81}$Br		&  3.803(-05)	&  0.3254	&  74669.264	&  7908	\\
HF						&  9.368(-05)	&  0.7187	&  18234.562	&  9418	\\
$^6$LiH				&  3.444(-05)	&  2.3149	&  6401.025	&  34303	\\
$^6$LiD				&  1.978(-05)	&  2.3094	&  7318.190	&  39029	\\
$^7$LiH				&  3.374(-05)	&  2.3143	&  7313.273	&  39170	\\
$^7$LiD				&  1.908(-05)	&  2.3087	&  8230.438	&  43868	\\
SiO						&  3.300(-06)	&  1.2190	&  40077.881	&  59555	\\
$^{73}$GeO		&  2.211(-06)	&  1.2915	&  81044.140	&  135174	\\
$^{74}$GeO		&  2.206(-06)	&  1.2915	&  81953.505	&  136690	\\
$^{6}$LiF			&  6.820(-06)	&  2.4895	&  22798.434	&  141300	\\
$^{7}$LiF			&  6.083(-06)	&  2.4885	&  23710.682	&  146835	\\
Ca$^{35}$Cl		&  6.931(-07)	&  1.6781	&  68295.833	&  192320	\\
Ca$^{37}$Cl		&  6.731(-07)	&  1.6781	&  70116.033	&  197446	\\
Li$^{35}$Cl		&  3.643(-06)	&  2.7935	&  37354.608	&  291510	\\
Li$^{37}$Cl		&  3.614(-06)	&  2.7935	&  39174.808	&  305714	\\
NaF						&  1.980(-06)	&  3.2089	&  38269.879	&  394078	\\
$^{7}$Li$^{79}$Br	&  2.518(-06)	&  2.4394	&  78324.361	&  466093	\\
$^{7}$Li$^{81}$Br	&  2.513(-06)	&  2.4394	&  80145.384	&  476930	\\
KF						&  1.270(-06)	&  3.3808	&  52829.231	&  603832	\\
Na$^{35}$Cl		&  9.899(-07)	&  3.5416	&  52826.053	&  662585	\\
$^{207}$PbO		&  1.396(-06)	&  1.8256	&  203225.448	&  677339	\\
$^{208}$PbO		&  1.396(-06)	&  1.8256	&  204137.580	&  680380	\\
Na$^{37}$Cl		&  9.900(-07)	&  3.5415	&  54646.253	&  685369	\\
$^{6}$LiI			&  2.338(-06)	&  2.9228	&  121148.795	&  1034936	\\
$^{7}$LiI			&  2.010(-06)	&  2.9228	&  122061.043	&  1042730	\\
K$^{35}$Cl						&  5.844(-07)	&  4.0404	&  67385.406	&  1100055	\\
K$^{37}$Cl						&  5.677(-07)	&  4.0403	&  69205.605	&  1129704	\\
SrO						&  1.535(-06)	&  3.5018	&  94699.539	&  1161235	\\
Na$^{79}$Br		&  6.870(-07)	&  3.5876	&  92883.558	&  1195526	\\
Na$^{81}$Br		&  6.832(-07)	&  3.5876	&  94704.581	&  1218965	\\
BaO						&  1.421(-06)	&  3.1295	&  140271.411	&  1373829	\\
NaI						&  5.353(-07)	&  3.6338	&  136620.240	&  1804044	\\
K$^{79}$Br		&  3.691(-07)	&  4.1816	&  107442.911	&  1878697	\\
K$^{81}$Br		&  3.661(-07)	&  4.1815	&  109263.934	&  1910503	\\
KI						&  2.768(-07)	&  4.2572	&  151179.593	&  2739934	\\
\end{longtable}
{\noindent}The wide variety of polar molecules available for trapping (once adequate experimental techniques
are developed) is one of several reasons for the intensity of recent interest in the field. 

At intermolecular separations much greater than the charge displacement, $r\gg s$, molecular interactions reduce to the familiar dipole-dipole
form 
\begin{equation}
  V(\vec{r})=\frac{d^{2}}{r^{3}}\left[\hat{d}_{1}\cdot\hat{d}_{2}-3(\hat{r}\cdot\hat{d}_{1})(\hat{r}\cdot\hat{d}_{2})\right]
\end{equation}
which, for a pair of identical molecules in the same field-aligned state, yields an equation of relative motion 
\begin{equation}
  \left[-\frac{\hbar^{2}}{2M}\nabla^{2}+\frac{\mu^{2}}{r^{3}}\left(1-3(\hat{r}\cdot\hat{z})^{2}\right)\right]\psi(\vec{r})=E_{{\rm rel}}\psi(\vec{r})
\label{eq:SchroedingerAU}
\end{equation}
where $\hat{z}$ is the field axis and $M$ is the reduced mass. 
As noted above, $\mu$ assumes its maximal value of $d$ at large fields, but reduces to zero as ${\cal E}\rightarrow 0$.

Equation (\ref{eq:SchroedingerAU}) is easily written in dimensionless form, {\it independent of $\mu$ and $M$}, 
\begin{equation}
\left[-\frac{1}{2}\nabla^{2}+\frac{1}{r^{3}}\left(1-3(\hat{r}\cdot\hat{z})^{2}\right)\right]\psi(\vec{r})
      =
     E_{{\rm rel}}\psi(\vec{r})\ \ \ .
\label{eq:SchroedingerDU}
\end{equation}
providing that all distances and energies are measured in dipole units (d.u.) 
\begin{equation}
\begin{array}{c}
  D=\mu^{2}M/\hbar^{2}\\
  E_{D}=\hbar^{6}/(M^{3}\mu^{4})
\end{array}\ \ .
\label{eq:DipoleScales}
\end{equation}
The emergence of these natural length and energy scales, previously noted by many authors~\cite{Scale}, is striking when one recognizes that for maximal 
$\mu$, $D\sim10^{2}-10^{6}$ times {\em larger} than typical molecular length scales, while $E_{D}\sim10^{-6}-10^{-12}$ times {\em smaller} 
than typical rotational level splittings. 
Dipole-dipole interactions dominate the long range part of the interaction in any polar system, suggesting that observable characteristics of polar
gases at sufficiently cold temperatures will be {\it universal}.  
Two types of long-range dipole-dipole effects can accordingly be identified: those that do not depend on the short-range boundary 
conditions and those that result from an interplay between short-range and long-range physics. 
In this work we model the short-range interaction with a spherically-symmetric "hard wall'' boundary  condition at $r=r_{0}\equiv R_{0}/D$.

\section{Adiabatic representation}

Expanding the solutions of Schr\"odinger's Eq.~(5) in terms of spherical harmonics seems a natural numerical approach with obvious 
advantages: the symmetries of the dipole-dipole interactions are clearly revealed in the spherical harmonic basis. 
The interaction conserves the parity of the state and the cylindrical symmetry allows for conservation of the quantum number $m$. 
The states of a given orbital quantum number $l$ are coupled only to the two states of $l\pm2$, which leads to a useful tri-diagonal structure of the potential matrix.

In a spherical harmonic representation, however, channel coupling is not small compared to the diagonal elements for the 
whole range of $l=0\ldots\infty$, and in the case of the s-wave, which is expected to contribute considerably at low energies, 
the coupling to the d-wave dominates the interaction. 
It is advantageous, therefore, to change the representation so that an efficient channel decoupling is achieved. 
For that purpose, we utilize the Born-Oppenheimer adiabatic representation, and diagonalize the angular part of the 
interaction together with the kinetic centrifugal terms.

For a given value of the interparticle distance $r$ we calculate the eigenstates of the following angular operator 
$H_{a}$
\begin{equation}
   [H_{a}(r)]_{ll'}=l(l+1)\delta_{ll'}+\frac{1}{r}\langle lm|1-3(\hat{r}\cdot\hat{z})^{2}|l'm\rangle\ \ .
\label{eq:adiaMatrix}
\end{equation}
As the potential matrix elements 
\begin{equation}
  \begin{array}{rcl}
    V_{ll'}^{(m)} &\equiv & \langle lm|1-3(\hat{r}\cdot\hat{z})^{2}|l'm\rangle  = \\ 
                         & = & (1-\frac{3}{2 l+1} 
     (\frac{(l-m) (l+m)}{2 l-1}+\frac{(l-m+1) (l+m+1)}{2 l+3})) \delta_{l,l'} - \\
     &  &
   -\frac{3}{2 l+3}   \sqrt{\frac{((l+1)^2-m^2) ((l+2)^2-m^2)}{(2 l+1) (2 l+5)}} \delta_{l,l'-2} \\
     &  &
   -\frac{3}{2 l-1}   \sqrt{\frac{((l-1)^2-m^2) (l^2-m^2)}{(2 l-3) (2 l+1)}} \delta_{l,l'+2}   
  \end{array}
  \label{eq:matrElement}
\end{equation}
depend on $m^2$, all the adiabatic states are doubly degenerate except $m=0$, which is non-degenerate.

Explicitly, we solve the following eigenvalue problem for each $r$
\begin{equation}
  H_{a}(r)\phi_{n}(\hat r;r)=v_{n}(r)\phi_{n}(\hat r;r)\ \ .
\label{eq:adiaEquation}
\end{equation}
As the off-diagonal part of the dipole-dipole interaction decreases at larger distances $r\rightarrow\infty$, all 
the adiabatic states $\phi_{n}$ approach  corresponding spherical harmonics and the potential curves $v_{n}/r^{2}$ 
approach the centrifugal barriers $l(l+1)/r^{2}$.
This allows us to classify the adiabatic states according to their long-distance behavior. 
For instance, in the bosonic case the lowest state approaches the $s$-wave, the next contributing adiabatic 
state has $d$-wave asymptotic behavior. 
We can therefore label the adiabatic potentials $v_{lm}(r)/r^2$ and the corresponding adiabatic eigenstates with 
quantum numbers $n\rightarrow l,m$ according to their long-distance angular behavior.

Constructing the adiabatic solutions of Eq.~(\ref{eq:adiaEquation}) by expansion in spherical harmonics,
\begin{equation}
\phi_{lm}(\hat r;r)= {\sum_{l^{\prime}\ge \vert m\vert} }^{\prime} \phi_{lm}^{l^{\prime}}(r) Y_{l^{\prime}m}(\hat r) 
\label{eq:10}
\end{equation}
we are now ready to expand the solutions of Eq.~(\ref{eq:SchroedingerDU}) in terms of the adiabatic states
\begin{equation}
\psi(\vec{r})=
  \sum_{l,m}F_{lm}(r)\phi_{lm}(\hat r; r) \ \ .
\end{equation}
(The prime on the sum in Eq.(\ref{eq:10}) indicates that only the terms with $l+l'=$even are included.)
The radial wavefunctions $F_{lm}(r)$ then satisfy the system of equations
\begin{equation}
  (-\frac{1}{2}\frac{d^{2}}{dr^{2}}+\frac{1}{r^{2}}v_{lm}(r)-E)F_{lm}(r)+\sum_{l'}(q_{ll'}^{(m)}(r)\frac{d}{dr}+p_{ll'}^{(m)}(r))F_{l'm}(r)=0 \ \ 
\label{eq:SchroedingerAD}
\end{equation}
The long-range behavior of $F_{lm}(r)$ coincides with the usual partial wave amplitudes, since the adiabatic components $\phi_{lm}(\hat r;r)$ approach 
the spherical harmonics $\phi_{lm}^{(l')}(r)\rightarrow\delta_{ll'}$ as $r\rightarrow\infty$. 
We have solved the coupled-channel Schr\"odinger equation~(\ref{eq:SchroedingerAD}) numerically, propagating solutions from the hard-sphere radius to asymptotic distances $r\sim500D$.
We employ quintic splines and orthogonal collocations~\cite{DeBooorSchwartz} for discretizing the Schr\"odinger equation and match the numerical solution with asymptotic boundary conditions at the right end of the interval.

Unlike the partial wave expansion, the coupling matrix elements $p$ and $q$ are not tridiagonal in $l$. 
For low energy calculations, however, the adiabatic representation is still advantageous, as one can see from Table~\ref{tab:Convergence}. 
The adiabatic representation clearly convergences rapidly, even in the vicinity of threshold resonances 
where the partial wave representation fails to converge with a given number of channels.

\begin{table}
\caption{\label{tab:Convergence} Convergence of the elastic cross section with respect to the number of channels for the incident energy of $E=5\times 10^{-4}$~d.u.~.
Results for adiabatic ({\em ad}) and partial wave ({\em p.w.}) representations are shown. The adiabatic representation demonstrates much faster convergence, especially in
the vicinity of a resonance.}
\begin{center}
\begin{tabular}{c|cc|cc|cc}
\parbox{10mm}{N channels}
&
  \multicolumn{2}{c}{\parbox{20mm}{ $r_0=0.01455$~d.u.}}
&
  \multicolumn{2}{c}{\parbox{20mm}{ $r_0=0.01457$~d.u.}}
&
  \multicolumn{2}{c}{\parbox{20mm}{ $r_0=0.01548$~d.u.}}
\\
& {\small \em ad.} & {\small \em p.w.} & {\small \em ad.} & {\small \em p.w.} & {\small \em ad.} & {\small \em p.w.}
\\
1  &12.39  &  0.005 & 11.27 &  0.005 & 0.45 & 0.006  \\
2  & 34.59 & 73.13 &  30.85 & 66.15 & 1.92 & 3.46  \\
4  & 3.14  & 16.00 & 169.476& 14.81  & 2.17  & 2.44 \\
8  & 3.30  & 2.25 & 144.78  & 1581.84  & 2.21 & 2.24 \\
16 & 3.30  & 3.10  & 144.79 & 121.58   & 2.21 & 2.25 \\
\end{tabular}
\end{center}
\end{table}

\begin{figure}[htb]
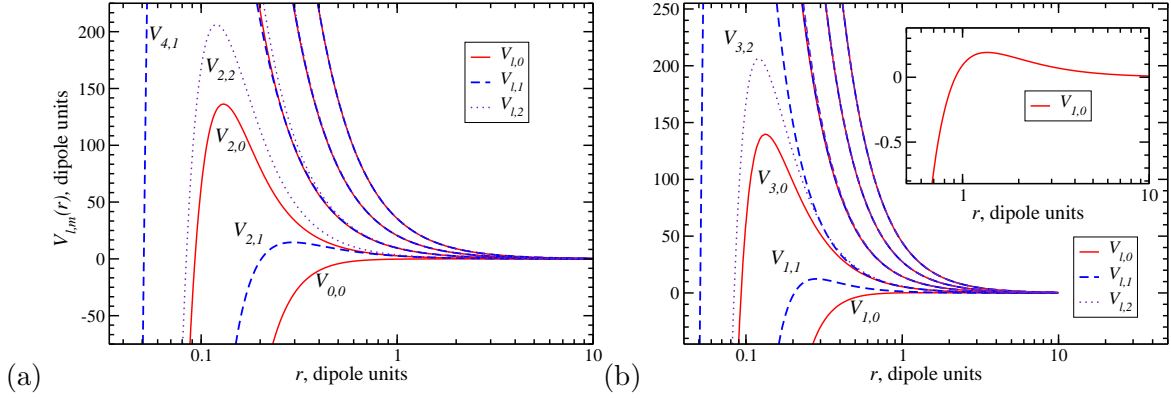

\begin{centering}
 (a) \includegraphics[clip=true,scale=0.3]{fig1a.eps}
 (b) \includegraphics[clip=true,scale=0.3]{fig1b.eps}
\end{centering}
\caption{\label{Fig:1} (Color online) Adiabatic potential curves for bosonic (a) and fermionic (b)
polar molecules. The curves are labeled by their asymptotic angular
momentum $\ell$ and by their exact magnetic quantum number $|m|$. Notice a small potential barrier in the lowest fermionic adiabatic potential $V_{1,0}$. }
\end{figure}

Neglecting the off-diagonal channel coupling in Eq.~(\ref{eq:SchroedingerAD}) yields an adiabatic approximation, in which
elementary one-dimensional potentials 
\begin{equation}
  V_{lm}(r)=\frac{v_{lm}(r)}{r^{2}}+p_{ll}^{(m)}(r)\ \ .
\end{equation}
afford a simple, intuitive picture of the scattering. 
These effective potentials are shown in Fig.~\ref{Fig:1} for bosonic (a) and fermionic (b) molecules. 
All the potentials have a very strong short-range attractive well. 
The long-range behavior in all the channels is $\frac{1}{r^2}$, excluding the lowest $V_{0,0}$. 
This lowest channel is attractive everywhere with a $\frac{1}{r^4}$ attractive tail.
All the other effective potentials are repulsive at large distances and have a single potential barrier in the transitional region between short-range attraction and long-range repulsion.
The barrier heights provide important additional energy scales for the system. 
As long as the collision energy does not exceed the barrier in the second adiabatic channel, the lowest channel strongly dominates and, therefore determines the character of the dipole-dipole scattering.
This serves to identify the "threshold regime" as corresponding to collision energies that are much smaller than the effective potential barriers.
Another important feature of the adiabatic potential curves is the presence of a small, $\sim 0.1$ d.u., barrier in the lowest fermionic channel. 
This barrier results in substantial differences between fermionic and bosonic threshold scattering.

\begin{figure}[htb]
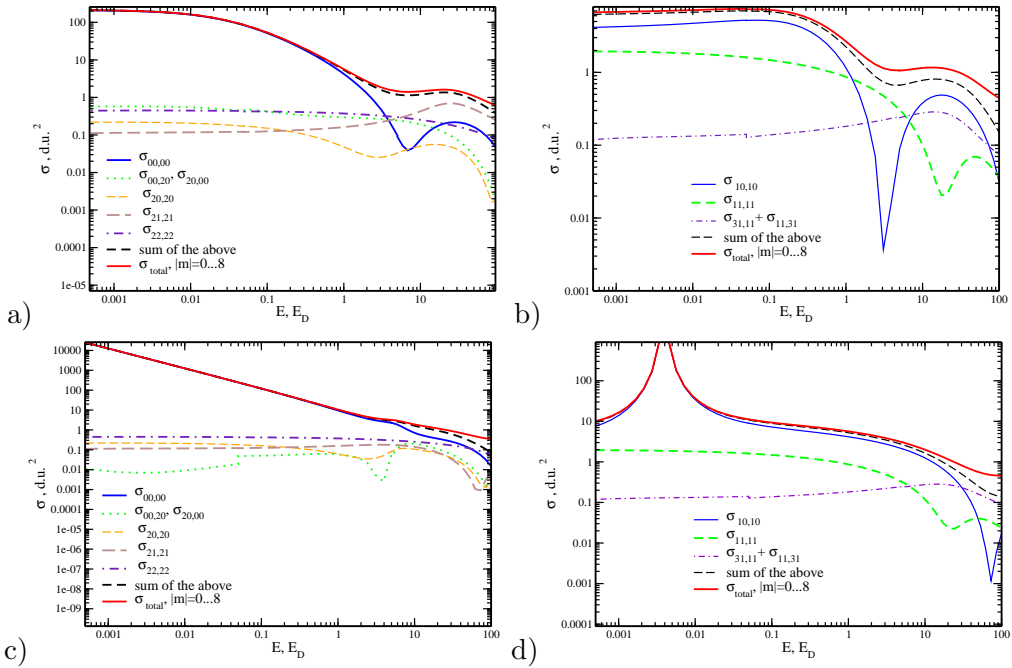

\begin{center}
\begin{tabular}{cc}
  a)\includegraphics[scale=0.25,clip=true]{fig2a.eps} & 
  b)\includegraphics[scale=0.25,clip=true]{fig2b.eps} \\ 
  c)\includegraphics[scale=0.25,clip=true]{fig2c.eps} & 
  d)\includegraphics[scale=0.25,clip=true]{fig2d.eps} \\ 
\end{tabular}
\caption{\label{Fig:CSEnergyDep} 
Energy dependence of the field-averaged scattering cross section.
a) Bosonic off-resonance scattering, $r_0=0.108$.
b) Fermionic off-resonance scattering, $r_0=0.035$.
c) Bosonic resonant scattering, $r_0=0.042873$
d) Fermionic resonant scattering, $r0=0.05693$
}
\end{center}
\end{figure}

\begin{figure}[htb]
\begin{center}
	\includegraphics[scale=0.50,clip=true]{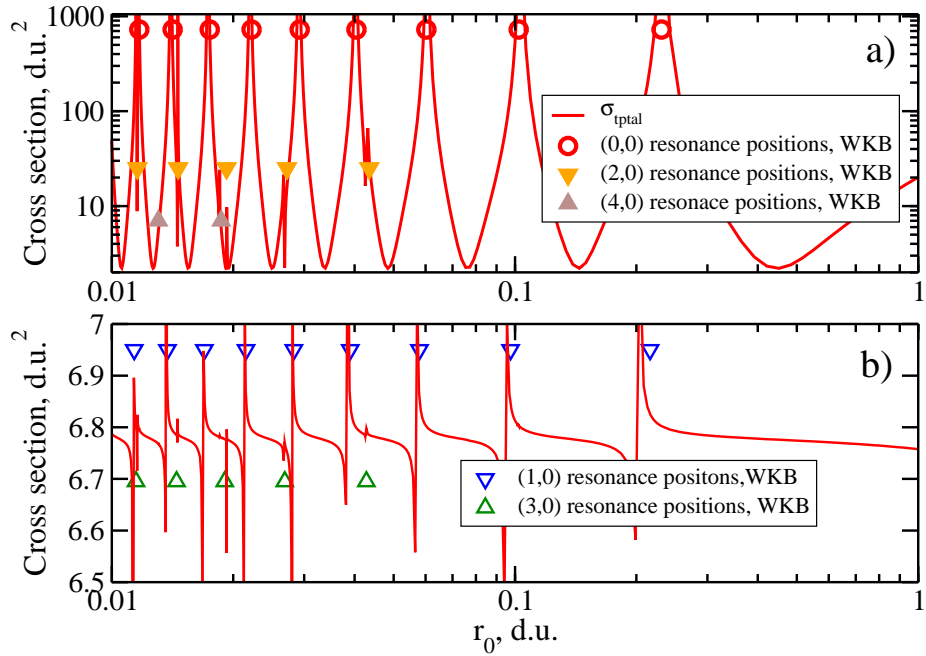}
\caption{\label{Fig:3} Near-threshold cross section as a function of the cut-off radius for bosons and fermions. Positions of the different $(l,m)$ resonances  predicted from WKB match numerical results.}
\end{center}
\end{figure}

Contributions of various channels to the total elastic cross-section for bosonic scattering (averaged over the electric-field direction), are shown in Fig.~\ref{Fig:CSEnergyDep}~a), for $r_{0}=0.0108$ dipole units.
The uncoupled $V_{0,0}$ and $V_{2,1}$ channels dominate bosonic scattering, while $V_{1,0}$ and $V_{1,1}$ dominate fermionic scattering, for energies 
below $\sim 130E_{D}$, as indicated above. 
Note, however, that detailed convergence of the cross-section (even at zero energy) requires several adiabatic channels. Both the magnitude of the threshold cross section and the degree of convergence vary with the hard-sphere radius, as they are sensitive to resonance formation in the inner wells of excited potential curves. 
This is illustrated in Fig.~\ref{Fig:3}, where the elastic cross section (averaged over the field direction) is plotted versus hard-sphere radius at the near-threshold energy, $E=5\times10^{-4}~E_{D}$. 
Also note that the threshold cross-section varies by 4-orders of magnitude, depending on the specific details of the short-range scattering.

The nature of these variations can be understood better after a closer look at Fig.~\ref{Fig:1}. 
Each of the adiabatic channels has a strong short-range attractive core which can support many bound states. 
When varying the short-range boundary conditions we effectively vary the number of bound states supported by each of the potentials.
Each time one of the adiabatic potentials looses a threshold bound state we see a strong variation in the total cross section.

Figure 2 therefore illustrates four distinct cases, depending on the presence of a near-threshold bound state and the system parity. 
The effect of a near-threshold resonance on the cross section energy dependence is different for bosons and fermions. 
Even though the threshold bound state strongly influences the magnitude of the cross section in the bosonic case, the energy dependence of the cross section remains qualitatively the same (Fig~\ref{Fig:CSEnergyDep}a,c). 
In the case of fermions, however, there is a striking difference between resonant and off-resonant cases (Fig~\ref{Fig:CSEnergyDep}b,d):  the shape resonance formed in the lowest adiabatic potential is narrow enough to produce a rapid variation of the cross section. 
In the non-resonant fermionic case (Fig~\ref{Fig:CSEnergyDep}b), when the collision energy approaches the barrier top of $\sim 0.1$d.u. the scattering cross section also 
grows, although not producing as distinctive a feature as in the resonant scenario (Fig~\ref{Fig:CSEnergyDep}d).
While bosonic threshold bound states produce a cross section enhancement in a broad range of the cut-off radii $r_0$, fermionic cross sections are strongly enhanced only
in very close vicinity of the resonant conditions (i.e. the resonances are clearly narrower).

The influence of  the short-range cut-off radius on the resonances' formation can be understood from simple semiclassical analysis, as discussed below.

\section{Semiclassical analysis}
One of the advantages of the adiabatic representation is that it provides a simple interpretation 
of the entire set of resonances shown in Fig.~\ref{Fig:3}. 

Let us look at the case of bosons first (Fig.~\ref{Fig:3}a). 
Consider the positions of wide peaks corresponding to a near-threshold bound state forming in the lowest adiabatic channel. 
Such bound states exist if the Bohr-Sommerfeld quantization condition (at $E=0$) is satisfied 
\[
  \int_{r_{0}}^{\infty}\sqrt{2\vert V_{0,0}(r)\vert}dr=n\pi+\phi_{0}\ \ .
\]
Although the semiclassical condition is not strictly applicable to the calculation of the position of the rightmost peak 
in Fig.~\ref{Fig:3}a (since the right turning point in the $V_{0,0}$ channel effectively lies at $r\rightarrow\infty$), we have 
introduced a single fitting parameter $\phi_{0}=0.0735\pi$ in order to reproduce the position of the rightmost peak correctly. 
Fitting the potential $V_{0,0}$ as $2 r^{3}V_{0,0} \approx-\frac{\alpha^2}{r+\beta}$ we get an explicit expression for the values 
of $r_{0}$ that support a near-threshold bound state 
\begin{equation}
  r_{0}^{(0,0)}(n)=\frac{4\alpha^{2}}{(4 \alpha+\beta (\pi n+\phi_{0}))(\pi n+\phi_{0})}\ \ .
\label{eq:sWaveRes}
\end{equation}
The universal dimensionless parameters $\alpha=0.9586$ and $\beta=0.265$ are obtained from fitting the adiabatic potential curve, and Eq. (\ref{eq:sWaveRes}) then 
gives the positions of the peaks in Fig.~\ref{Fig:3}a up to 3 significant figures.

Similarly, the semiclassical description is useful for understanding positions of other resonances that occur in higher adiabatic channels. 
All the resonances in the model system can be classified according to the asymptotic behaviour of the effective potential that produces 
the resonance, i.e. the corresponding $(l,m)$ quantum numbers, and the number of bound states supported by the corresponding channel potential. 
If nonadiabatic coupling could be neglected, the resonances would have zero-width when approaching threshold because of the large 
dipole-modified centrifugal barrier in Fig.~\ref{Fig:1}. 
In fact, they dominantly decay into the lowest $m=0$ adiabatic channel through short-range nonadiabatic coupling and 
show up as Feshbach resonances in the total elastic cross section. 
Their positions can be found directly from the usual Bohr-Sommerfeld quantization procedure with no fitting parameters. 
Although it is more difficult to develop a simple fit to the higher adiabatic potentials, an expression similar 
to (\ref{eq:sWaveRes}) provides an excellent fit to the positions of the other resonances obtained from analyzing the semiclassical picture numerically 
\begin{equation}
  r_{0}^{(l,m)}(n)=\frac{1}{a_0+a_1 n+a_2 n^{2}}\ \ .
\label{eq:dWaveRes}
\end{equation}
We have performed such fitting for the dominant channels contributing to the low-energy scattering and the results are presented in Table~\ref{tab:ResParam}.
\begin{table}
\caption{\label{tab:ResParam} Fitting parameters for resonant short-range cut-off positions $r_0$ for the channels dominating at low energy.}
\begin{center}
\begin{tabular}{c|ccc}
$(l,m)$ & $a_0$ & $a_1$ & $a_2$ \\
\hline
\multicolumn{4}{c}{Bosons} \\
\hline
	$(0,0)$			&	0.2447		&		3.3817		&		0.711496		\\
	$(2,0)$			&	8.17602	  &		13.5762		&		0.622266		\\
	$(4, 0)$ &	 	31.6553 &		21.2334 	&		0.63889	\\
\hline
\multicolumn{4}{c}{Fermions} \\
\hline
	$(1,0)$	&		0.3635		&		3.567		&		0.6844		\\
	$(3,0)$			&			10.873	&		11.809		&		0.68644		\\
	$(5,0)$			&			32.579	&		20.671		&		0.63113		\\
\end{tabular}
\end{center}
\end{table}

The case of fermions is very similar to the bosonic one, but the potential barrier in the lowest adiabatic channel makes the overall picture quite different: 
It makes the resonances formed in the lowest adiabatic channel comparatively narrow in both the energy and cut-off radius domains.

\section{Scattering anisotropy and threshold resonances.}
\begin{figure}[htb]
\begin{tabular}{cccc}
\multicolumn{4}{c}{
\begin{picture}(0,0)(0,0)
  \put(-50,120){ a)}
\end{picture}
\includegraphics[height=1.5in,clip=true]{profileBosons.eps} } 
\\
$\sigma$ & $d\sigma(\hat{k})$, $\theta=0$  & $d\sigma(\hat{k})$, $\theta=\pi/2$  & $d\sigma(\hat{k})$, $\cos^2\theta=1/3$  \\
\multicolumn{4}{l}{b) $r_0=0.014556$ } \\
\includegraphics[width=1in,clip=true]{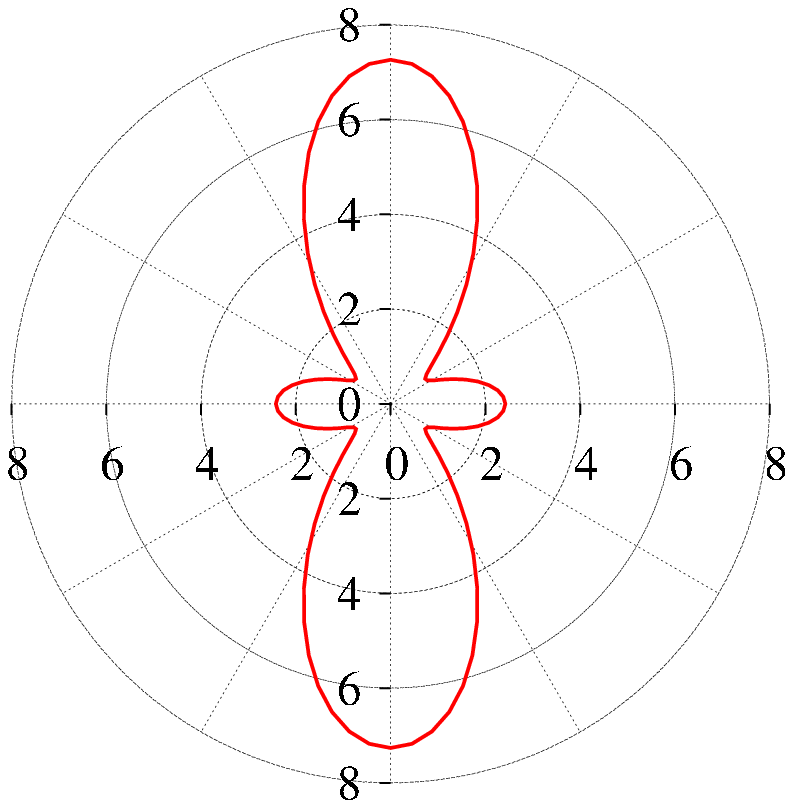} &
\includegraphics[width=1.3in,clip=true]{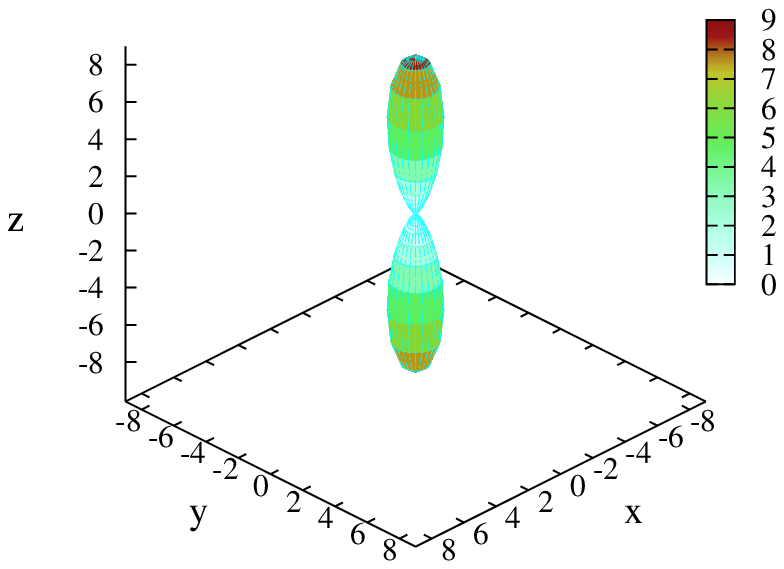} &
\includegraphics[width=1.3in,clip=true]{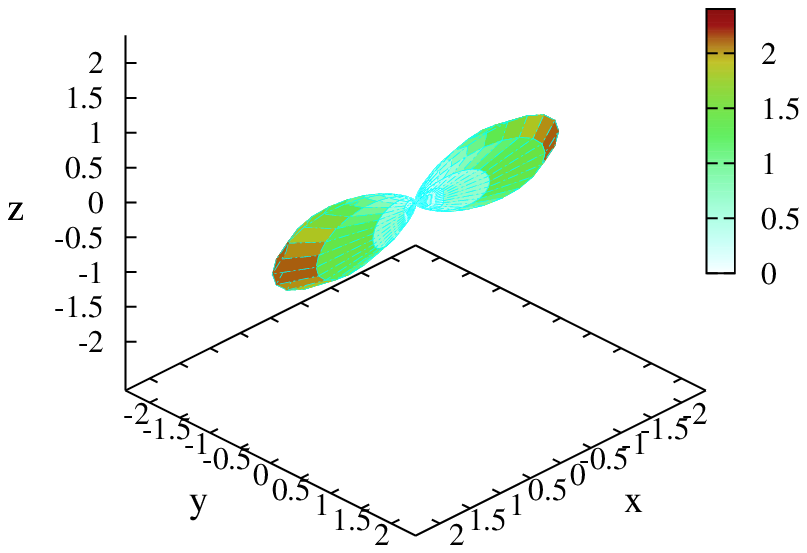} &
\includegraphics[width=1.3in,clip=true]{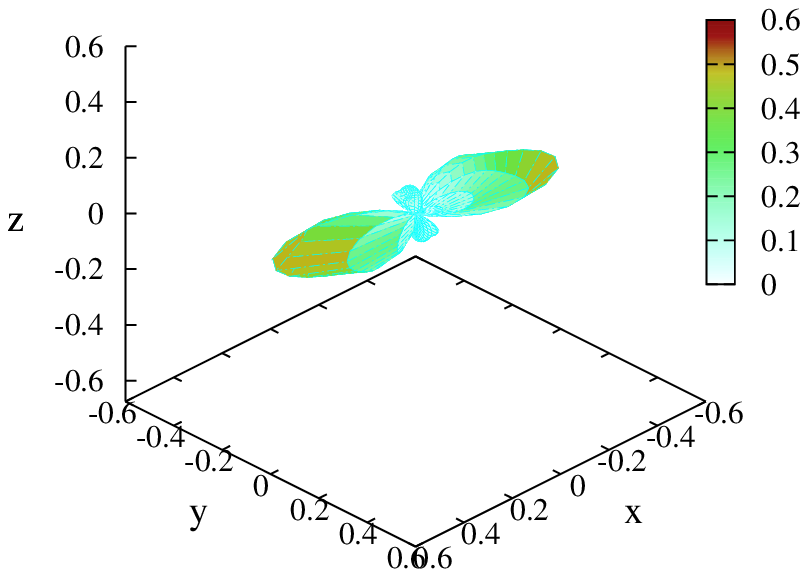} \\
\multicolumn{4}{l}{ c) $r_0=0.014572$ } \\
\includegraphics[width=1in,clip=true]{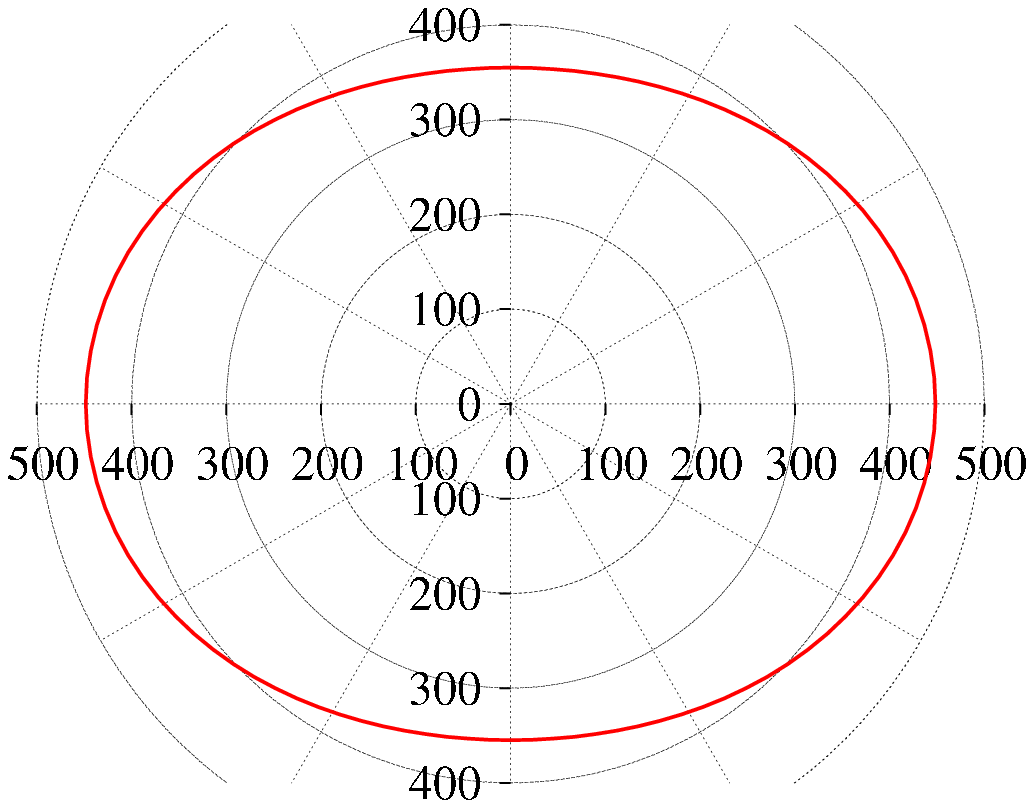} &
\includegraphics[width=1.3in,clip=true]{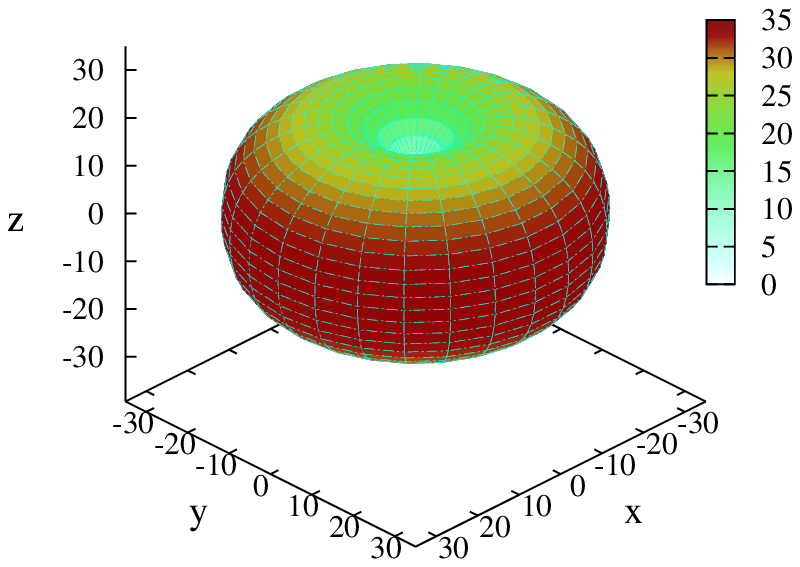} &
\includegraphics[width=1.3in,clip=true]{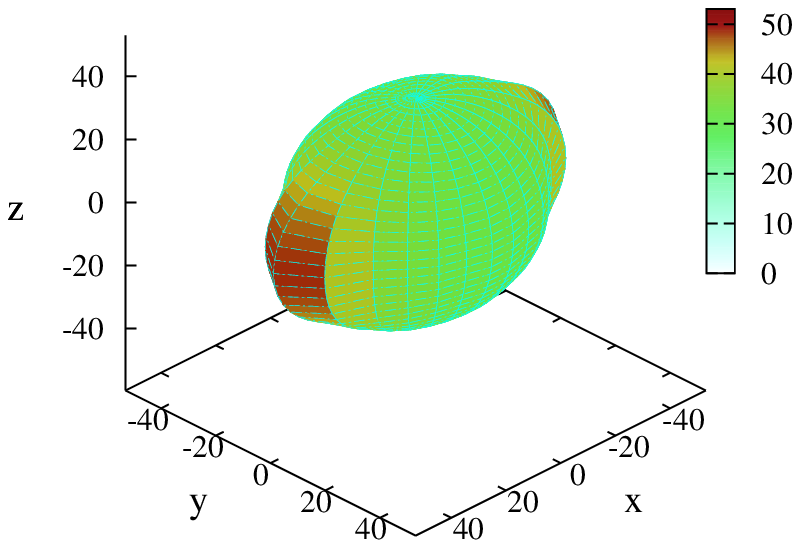} &
\includegraphics[width=1.3in,clip=true]{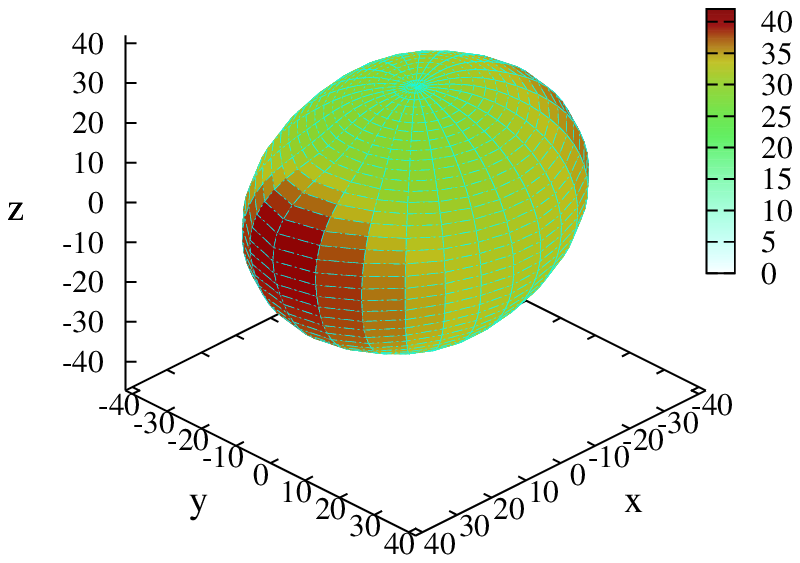} \\
\multicolumn{4}{l}{ d) $r_0=0.014576$ } \\
\includegraphics[width=1in,clip=true]{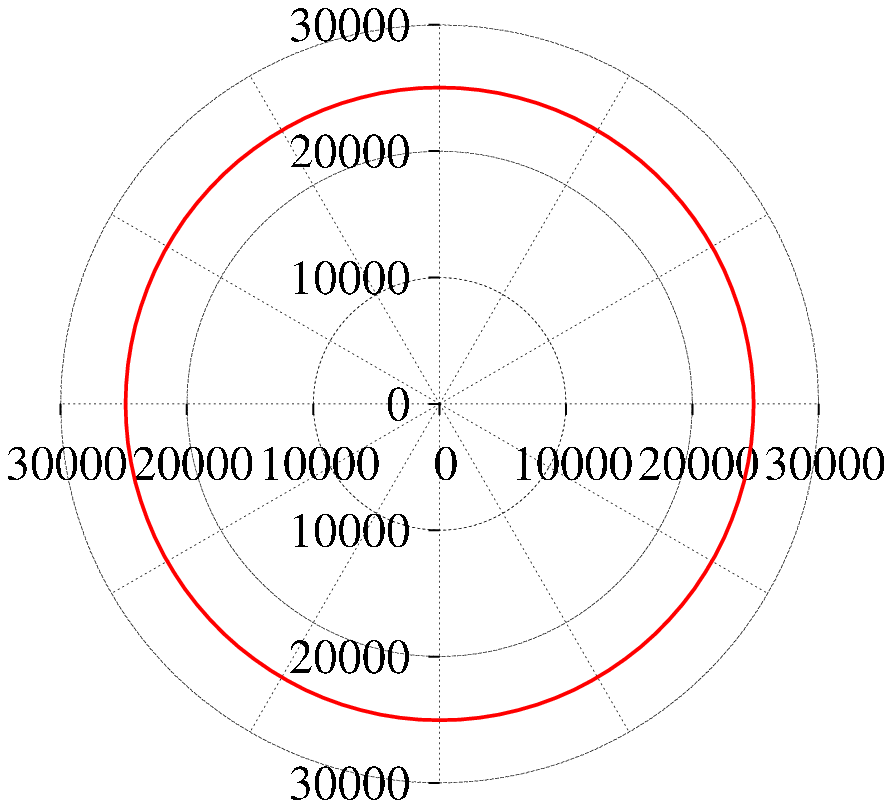} &
\includegraphics[width=1.3in,clip=true]{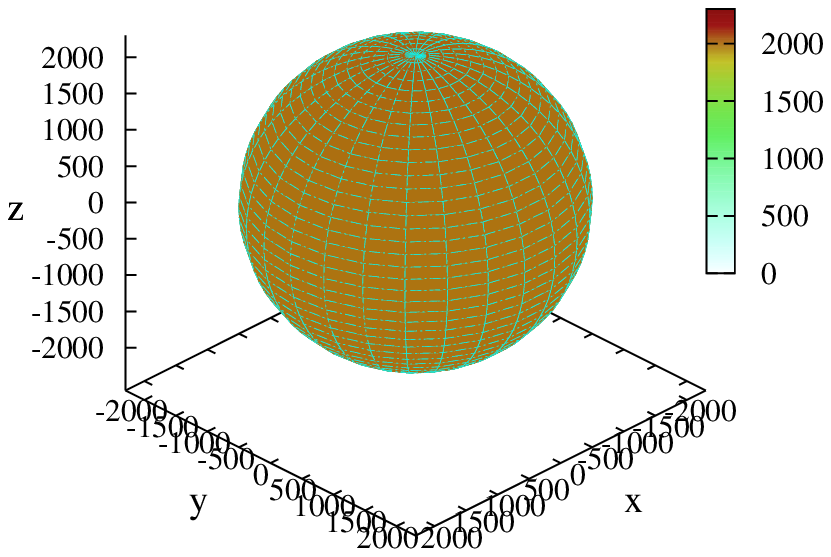} &
\includegraphics[width=1.3in,clip=true]{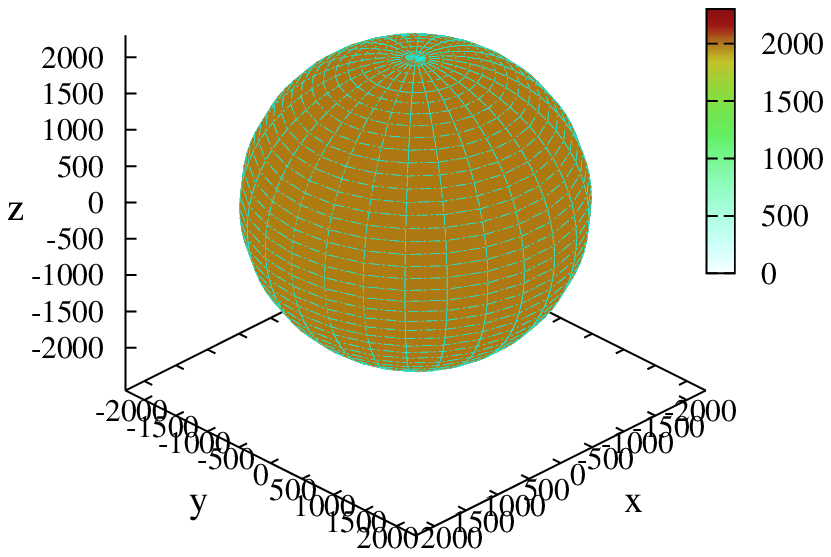} &
\includegraphics[width=1.3in,clip=true]{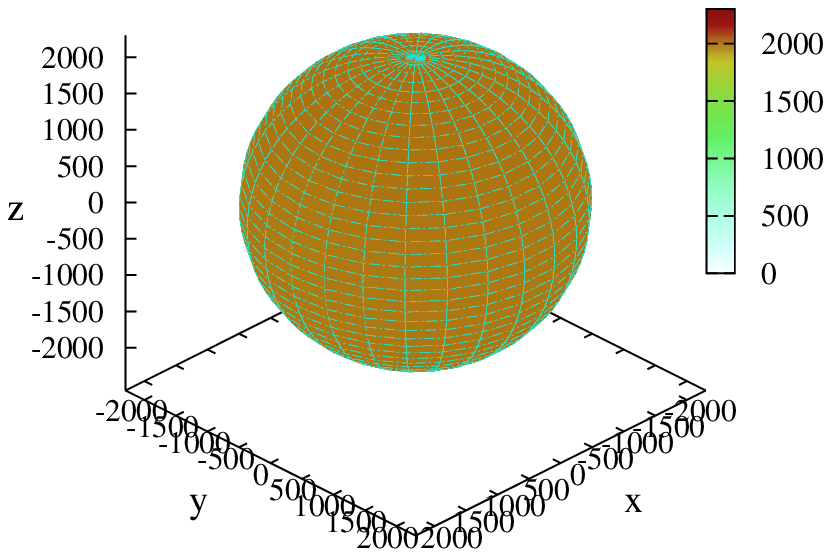} \\
\multicolumn{4}{l}{ e) $r_0=0.014579$ } \\
\includegraphics[width=1in,clip=true]{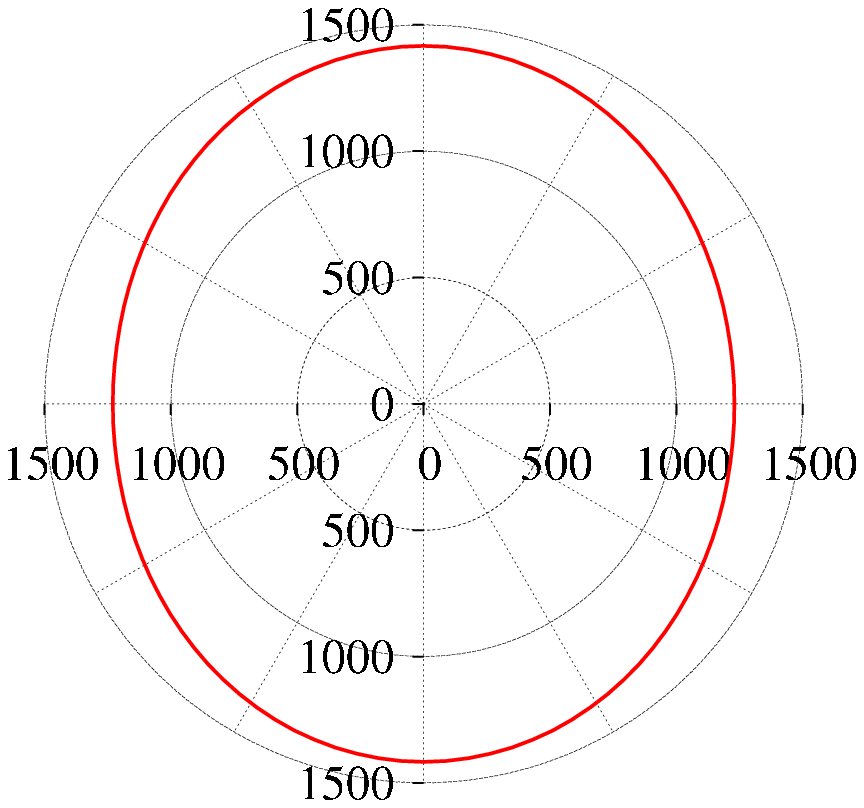} &
\includegraphics[width=1.3in,clip=true]{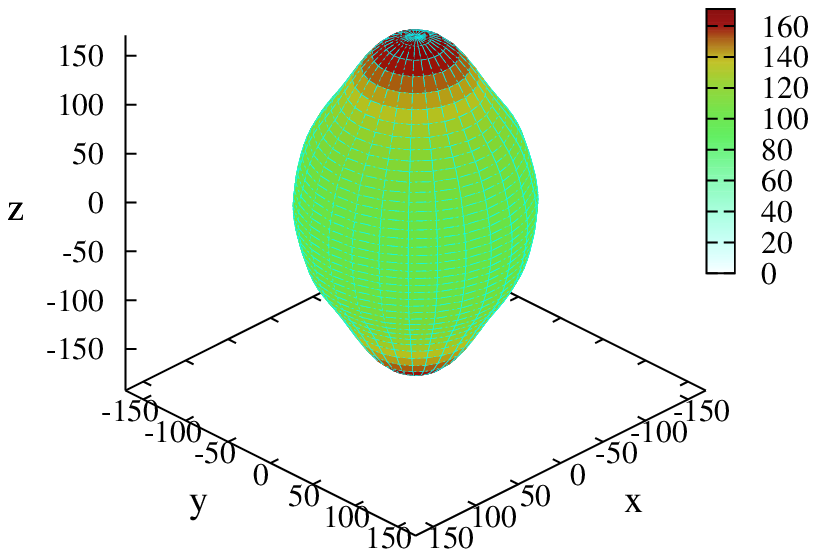} &
\includegraphics[width=1.3in,clip=true]{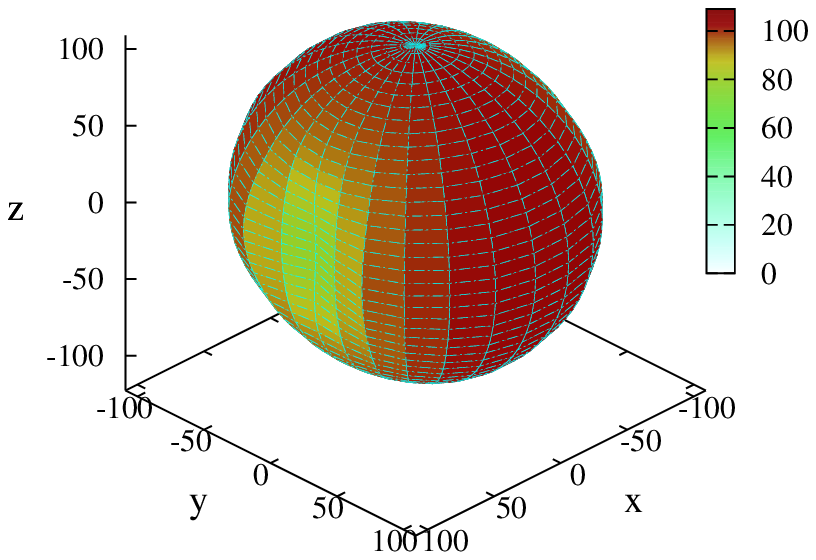} &
\includegraphics[width=1.3in,clip=true]{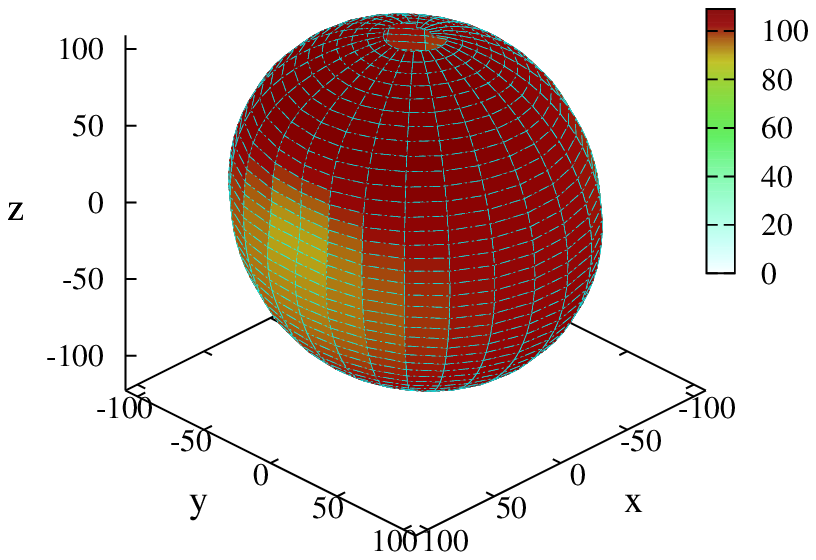} \\
\end{tabular}
\caption{\label{fig:BosonsAnis} a) Total cross section for two bosons averaged over field direction in the vicinity of the $(0,0)$ (diamond) and $(2,0)$ (circles) resonances. b)-e) Total and differential cross section dependence on the angle $\theta=\cos^{-1} \hat{k}\cdot \hat{z}$ between the incoming wave and the field direction for bosons. Both the total cross section and the differential cross section demonstrate strong dependency on the short-range physics. The dipoles are polarized along the $\hat z$ direction, $xz$ defines the scattering plane.}
\end{figure}

Unlike central potentials, the threshold scattering of dipoles can be strongly anisotropic even in the case of identical bosons. 
Although the $(0,0)$ channel strongly dominates scattering at low energies, the degree of this domination depends on short-range boundary conditions. 
For instance, when the total cross-section is minimal, the $(0,0)$ channel does not contribute to scattering, and we can expect the cross section to be strongly anisotropic. 
This is illustrated in Fig. (\ref{fig:BosonsAnis}b). At the minimum of the total cross section in the vicinity of the $(2,0)$ resonance (Fig. \ref{fig:BosonsAnis}a) 
the isotropic component of the cross section practically vanishes, and we observe a strong anisotropy both in total cross section i
(as a function of the incoming wave direction $\theta$) and in differential cross section 
(as a function of the incoming wave direction and the scattering angle). In the resonant case, however, when the total cross section is 
maximal (Fig.~\ref{fig:BosonsAnis}d), both total and differential cross sections become essentially isotropic with strong domination of the $(0,0)$ channel (s-wave scattering).

\begin{figure}
\begin{tabular}{cccc}
\multicolumn{4}{c}{
\begin{picture}(0,0)(0,0)
  \put(-50,120){ a)}
\end{picture}
\includegraphics[height=1.5in,clip=true]{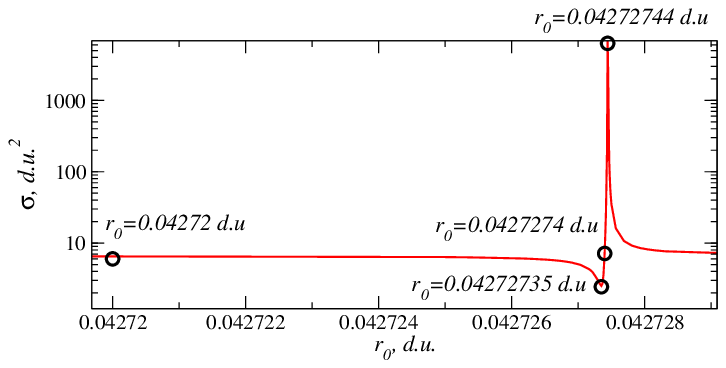} } \\
$\sigma$ & $d\sigma(\hat{k})$, $\theta=0$  & $d\sigma(\hat{k})$, $\theta=\pi/2$  & $d\sigma(\hat{k})$, $\cos^2\theta=1/3$  \\
\multicolumn{4}{l}{b)  $r_0=0.04272$ } \\
\includegraphics[width=1in,clip=true]{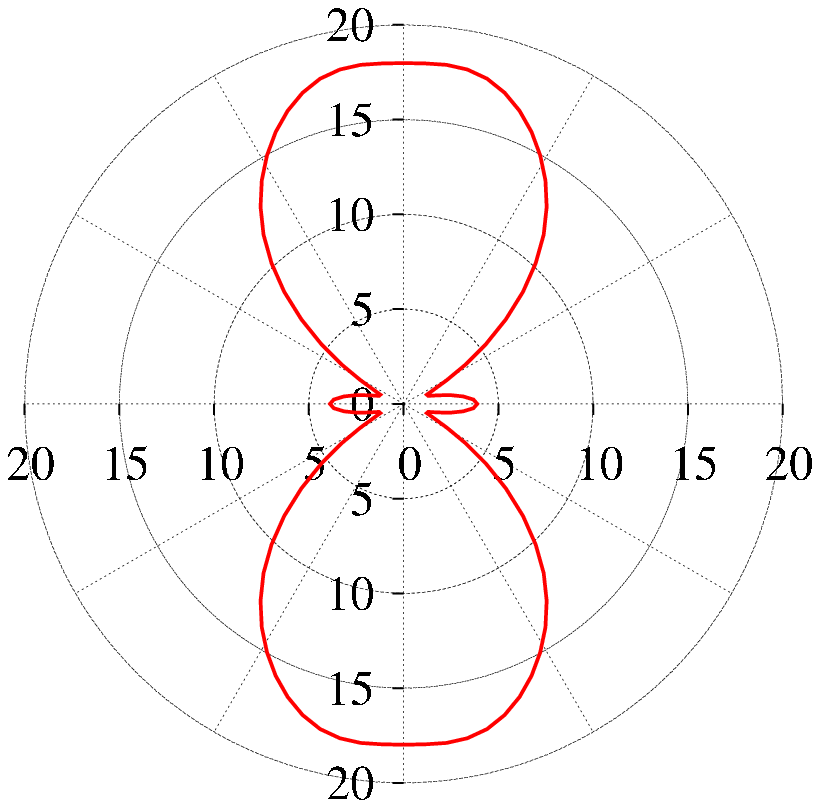} &
\includegraphics[width=1.3in,clip=true]{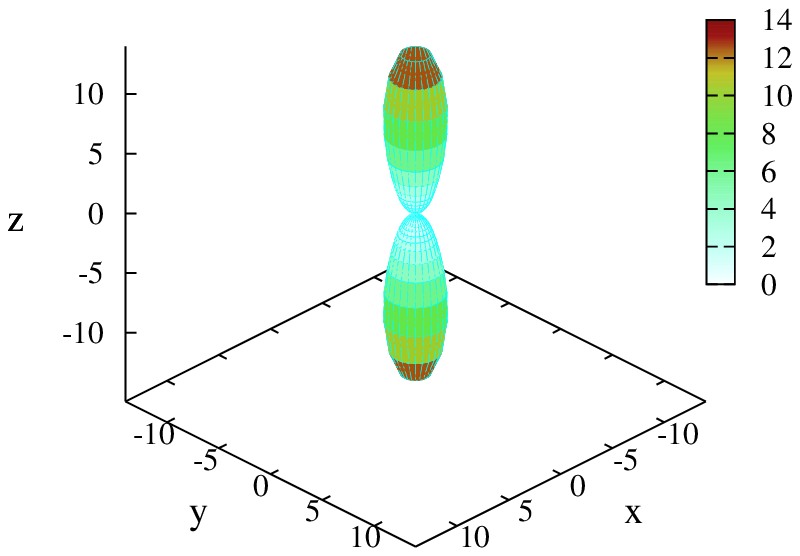} &
\includegraphics[width=1.3in,clip=true]{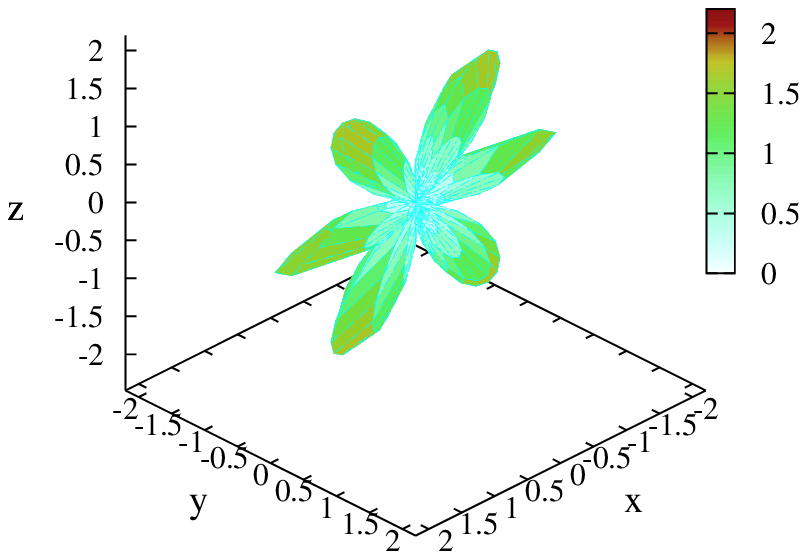} &
\includegraphics[width=1.3in,clip=true]{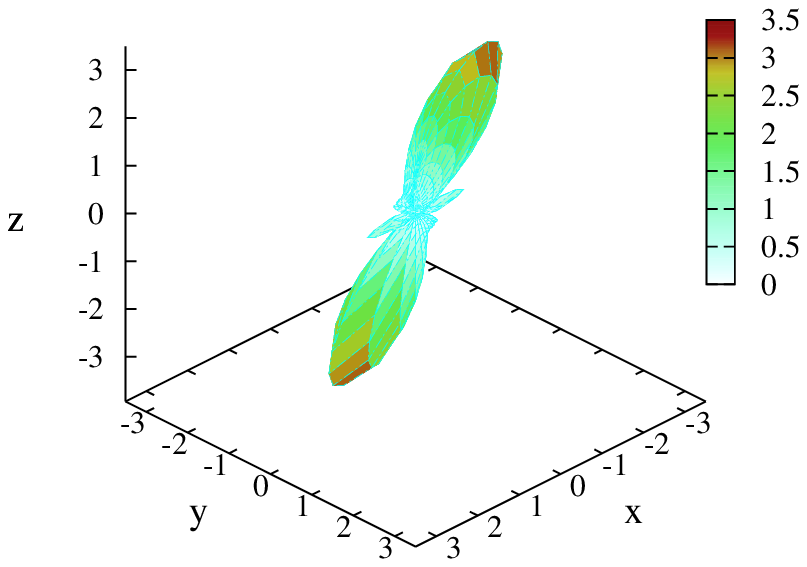} \\
\multicolumn{4}{l}{c) $r_0=0.04272735$ } \\
\includegraphics[width=1in,clip=true]{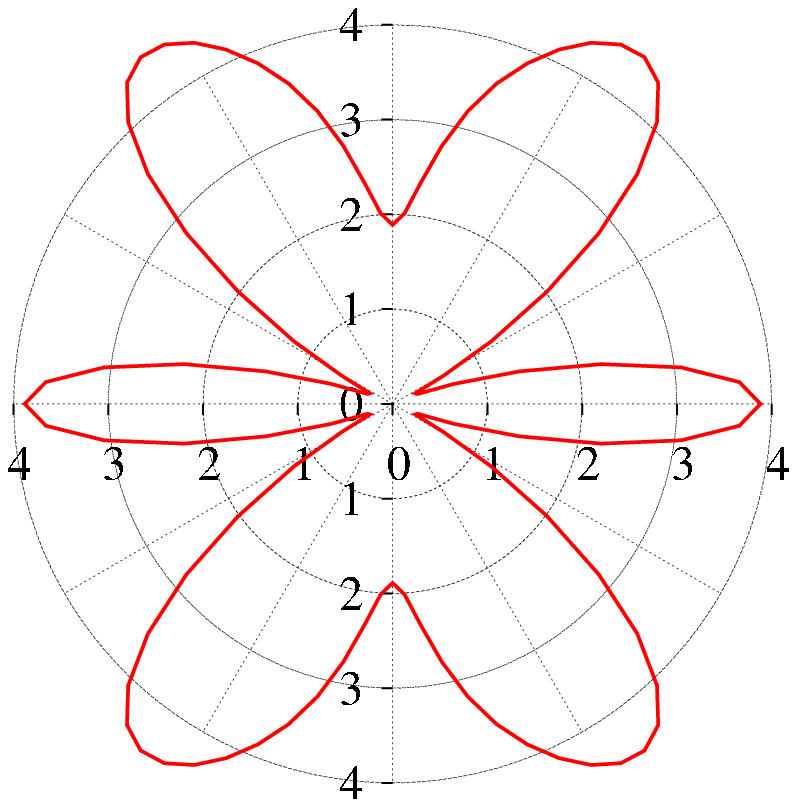} &
\includegraphics[width=1.3in,clip=true]{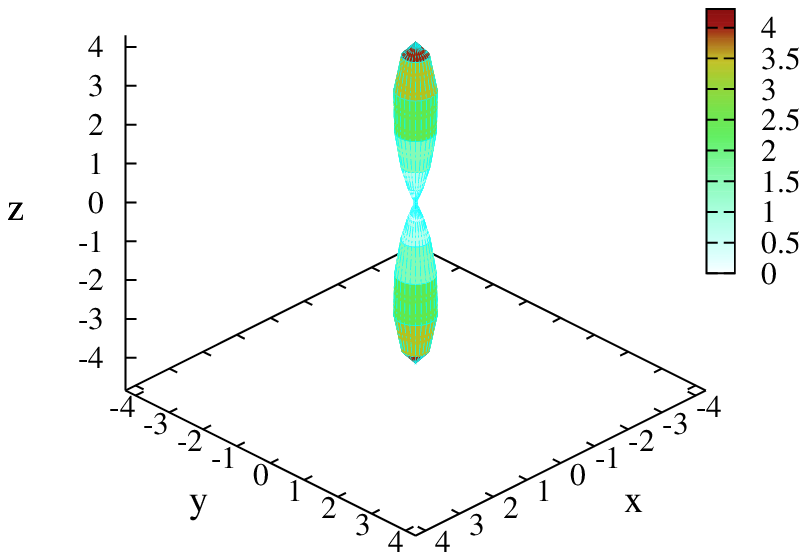} &
\includegraphics[width=1.3in,clip=true]{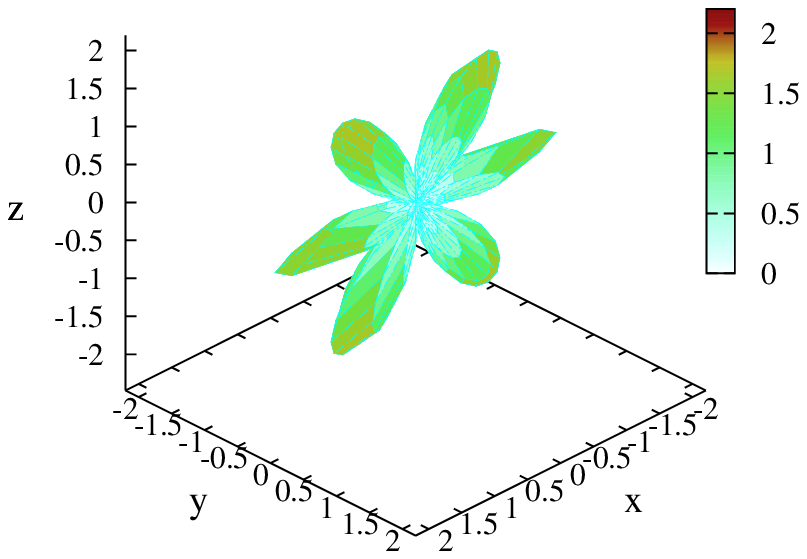} &
\includegraphics[width=1.3in,clip=true]{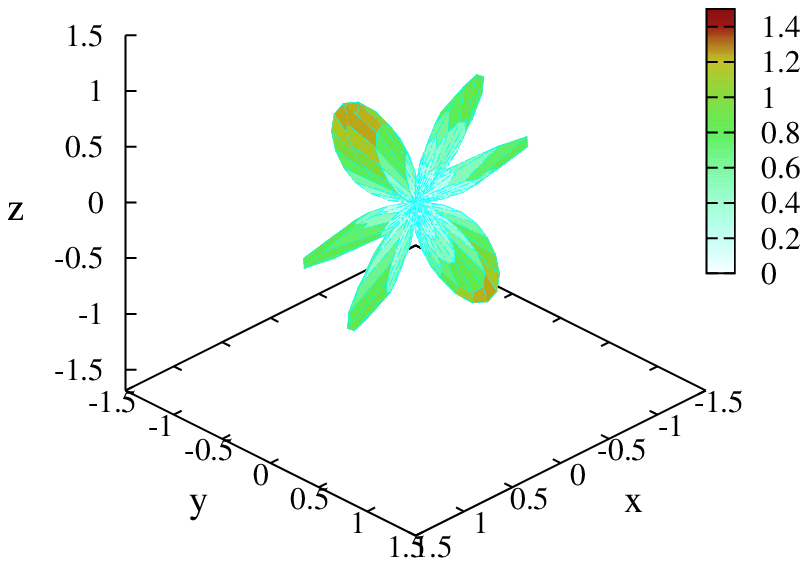} \\
\multicolumn{4}{l}{d) $r_0=0.0427274$ } \\
\includegraphics[width=1in,clip=true]{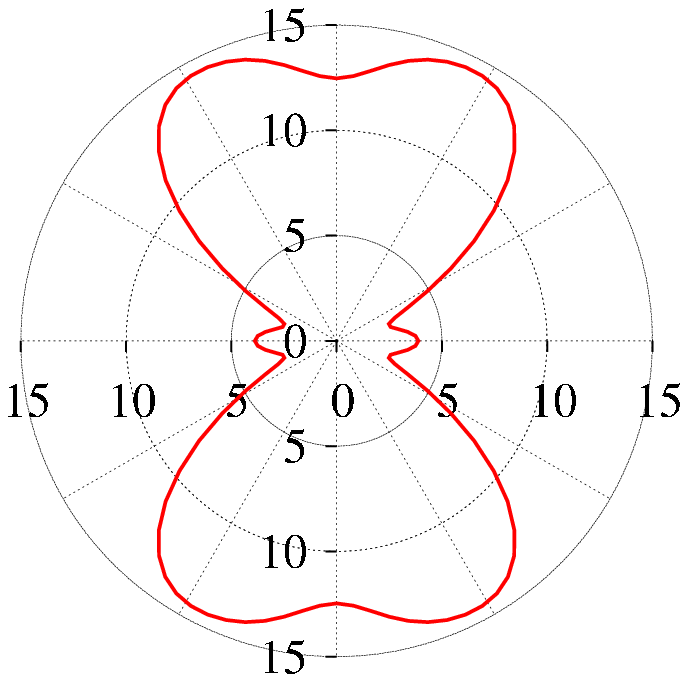} &
\includegraphics[width=1.3in,clip=true]{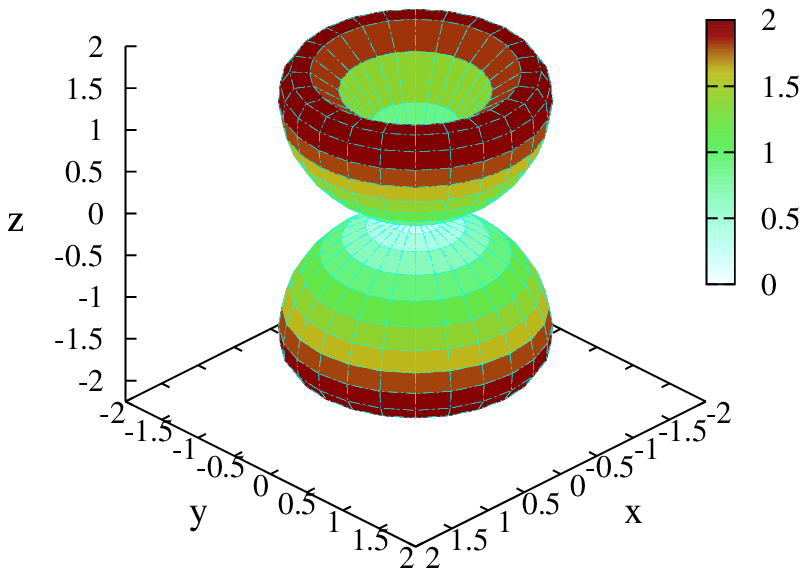} &
\includegraphics[width=1.3in,clip=true]{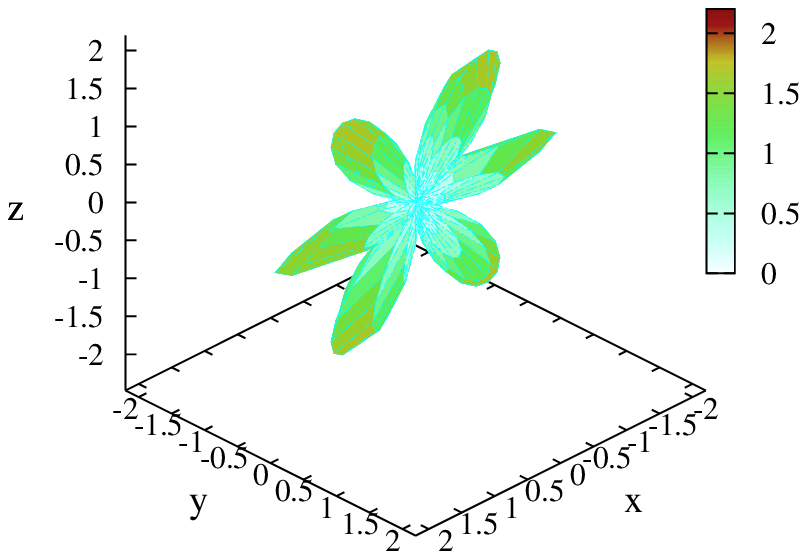} &
\includegraphics[width=1.3in,clip=true]{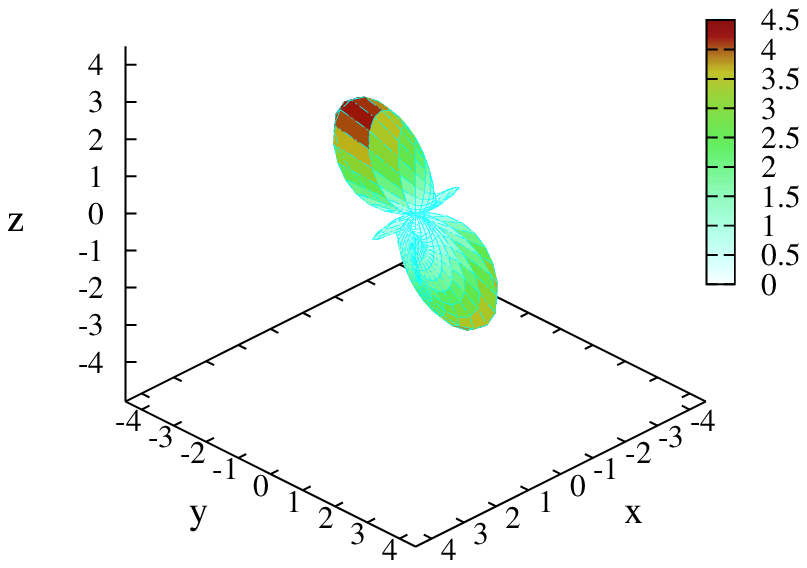} \\
\multicolumn{4}{c}{e) $r_0=0.0427444$ } \\
\includegraphics[width=1in,clip=true]{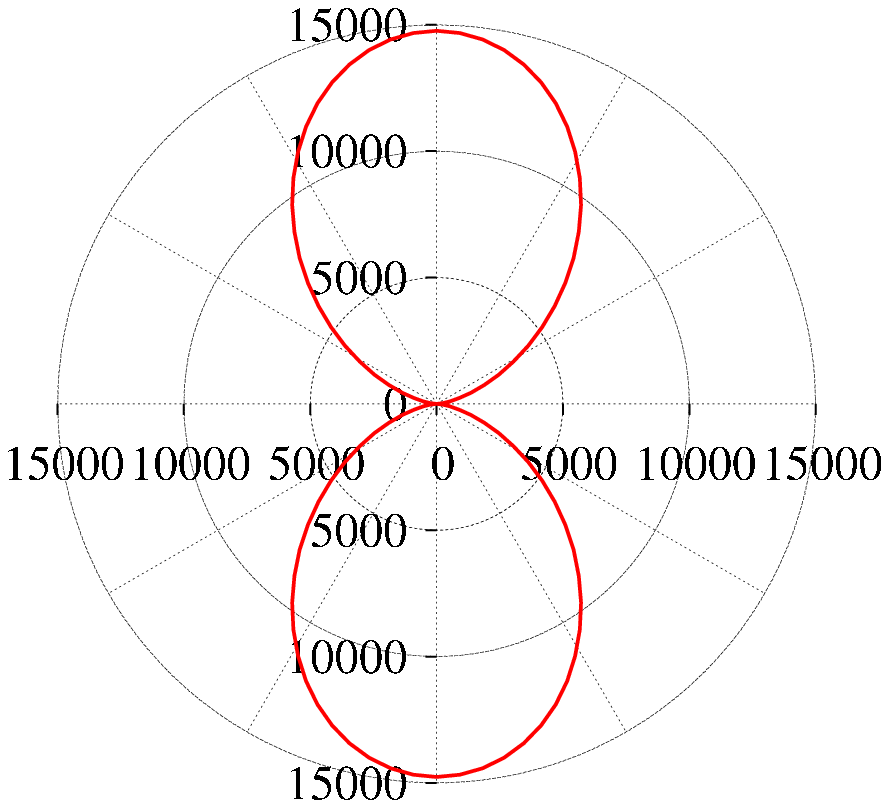} &
\includegraphics[width=1.3in,clip=true]{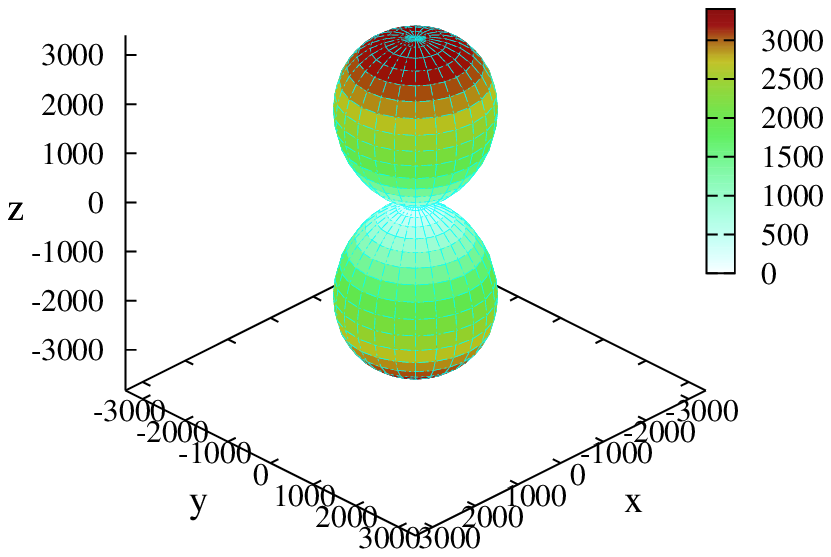} &
\includegraphics[width=1.3in,clip=true]{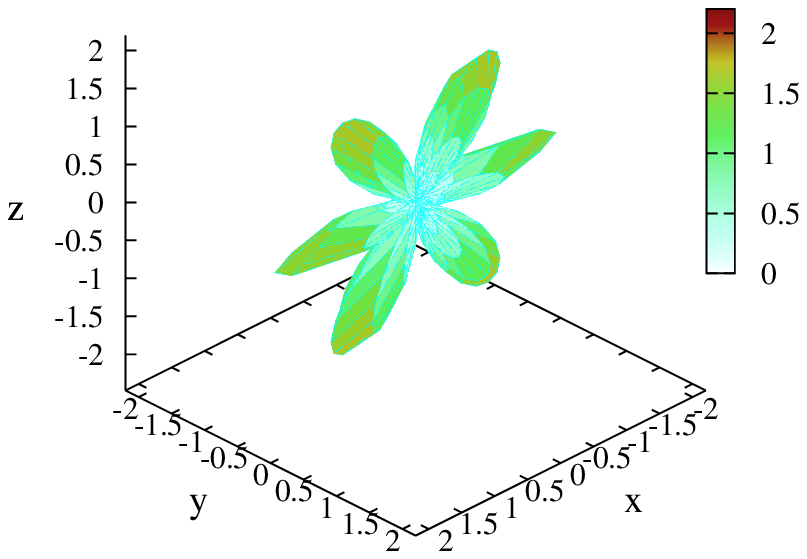} &
\includegraphics[width=1.3in,clip=true]{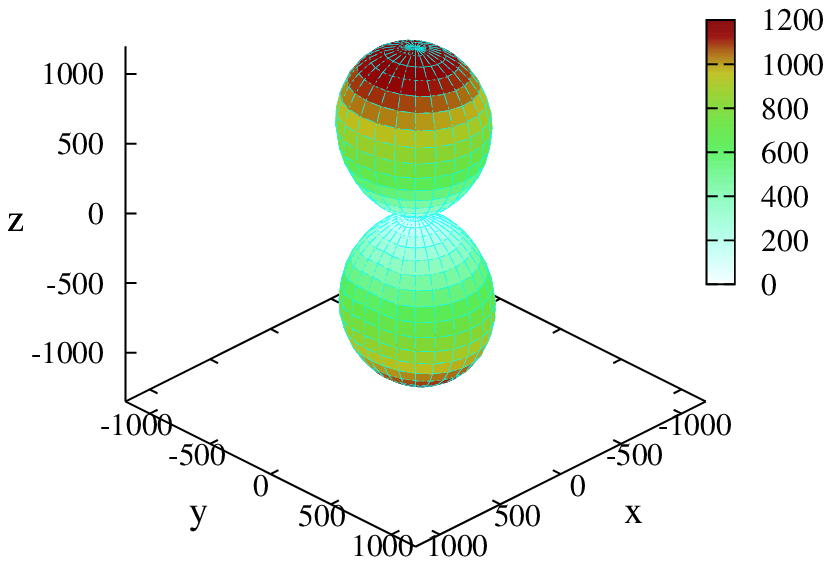} \\
\end{tabular}
\caption{\label{fig:FermionsAnis} Total and differential cross section dependence on the angle  $\theta=\cos^{-1} \hat{k}\cdot \hat{z}$ between the incoming wave and the field direction for fermions. Both the total cross section and the differential cross section demonstrate strong dependency on the short-range physics. Scattering perpendicular to the field is not sensitive to the short-range conditions, however.}
\end{figure}

A similar situation occurs for fermions. The angular distribution (which is always anisotropic) is dominated by the rapidly varying $(1,0)$ adiabatic channel, 
and the rest of the channels that are not that sensitive to the short-range physics. As in the case of bosons, we observe a rapid variation of the 
angular distributions in the vicinity of a $(3,0)$ resonance (Fig.~\ref{fig:FermionsAnis}). An interesting feature of the fermionic scattering, however, 
is the presence of a special direction $\theta=\pi/2$, i.e. collisions perpendicular to the field. Since the contribution of the $(1,0)$ component 
vanishes in this special direction, scattering perpendicular to the polarizing field is not sensitive to the details of the short-range interaction. 

\section{Some observability notes}
Sensitivity to $r_{0}$ indicates a corresponding sensitivity to short-range physics, and therefore, for a given electric field, to the specific molecular species, and some molecules can be expected to be dominated by either resonant or non-resonant adiabatic channels. 
The dipole moment induced by an external electric field, however, varies with the field until the dipoles are completely polarized. 
This makes it possible to observe, in principle, the series of resonances which we described in previous section. 

Assuming the electric field is large enough to polarize the molecule, but small enough not to perturb the short-range wave function, 
the only part of the intermolecule interaction affected by the field would be the dipole-dipole interaction. 
Changing the induced dipole $\mu$ by tuning the electric field we can effectively manage the dipole scales. 
The elastic scattering cross section $\sigma$ will, therefore, scale with the field ${\cal E}$ as 
\[\sigma =D({\cal E})^2 \sigma(E/E_D({\cal E}))\] where $D({\cal E})$ and $E_D({\cal E})$ are given by Eq.~\ref{eq:DipoleScales}. 
As aforementioned, the explicit dependence of the induced dipole moment on the field ${\cal E}$ depends on the molecular state.

As soon as the induced dipole becomes big enough for the dipole length to exceed the short-range scale about 7 times ($R_0/D\approx 0.14$), the first bound state is formed in the $V_{0,0}$ potential and the scattering cross section peaks. 
This is the first peak in the series (\ref{eq:sWaveRes}). 
This first peak would allow an experimentalist to identify the unknown empirical parameter $R_0$ and estimate positions of the subsequent peaks of the series (\ref{eq:sWaveRes}) from the condition $r_{0}^{(0,0)}(1)=\frac{R_0}{D({\cal E})}$. 
The number of peaks that can be potentially observed for particular molecular species is limited by their intrinsic dipole moment: as $\mu\leq d$, there is a maximal possible dipole scale $D_{max}$, and, thus, the number of peaks $n_{max}$ can be estimated from the condition $r_{0}^{(l,0)}(n_{max})\geq \frac{R_0}{D_{max}}$. 

\begin{figure}[htb]
\begin{center}
	\includegraphics[scale=0.50,clip=true]{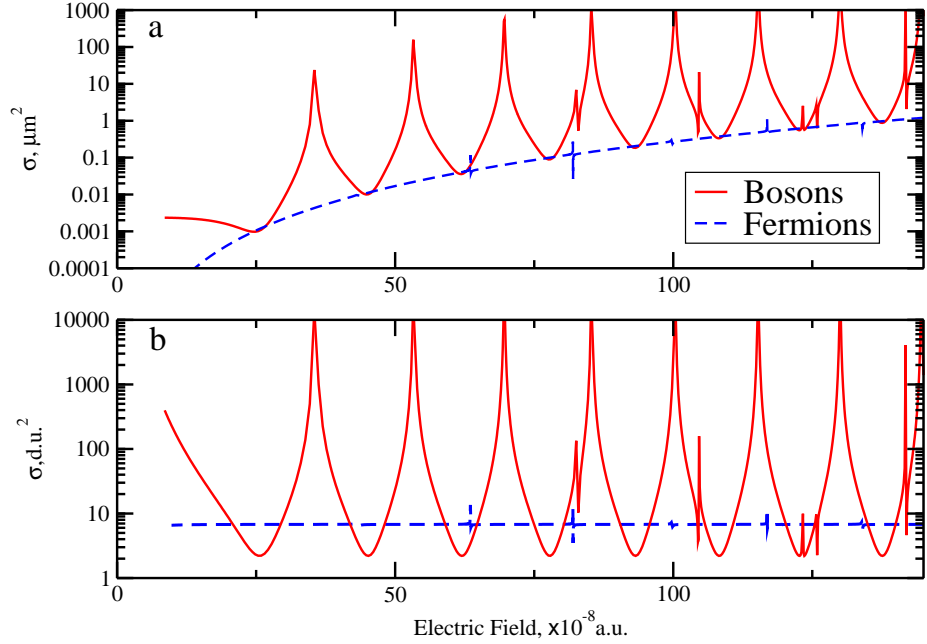} 
\caption{\label{fig:FieldDependence}Electric field dependence of the near-threshold $^{6,7}$LiF elastic scattering cross section, in both (a) absolute and (b) dipole units. 
The minima of the cross section in the upper figure rise rapidly with the field as a result of rapidly growing dipole length. The fermionic cross section exhibits much narrower resonant variations of the cross section while demonstrating the same ${\cal E}^4$ growing trend.}
\end{center}
\end{figure}

Using the parameters for $^{6,7}$LiF molecules given in Table~\ref{Tab:SigmaMol}  as an example and deliberately choosing $R_0=80$~a.u. as a short-range cutoff parameter, we show the direction-averaged total cross section as a function of the external polarizing field in Fig.~\ref{fig:FieldDependence}. 
For weak fields the induced dipole moment is small and, thus, the dipole units are comparable to the short-range molecular scale. 

Another interesting and universal feature of the dipole-dipole scattering cross section as a function of electric field is the trend of going up with the field.
This trend can be seen much more clearly in the case of fermions which exhibits much narrower resonant variations of the cross section.

\section{Summary}
An adiabatic representation is shown to be useful in numerical calculations of the ultra-low energy dipole-dipole scattering as well as for the classification of 
resonances that emerge in such systems. For both bose and fermi collision partners, the major resonant phenomena are determined by the $m=0$ channels, 
especially the lowest channel in the $m=0$ set. The resonant states formed result from the long-range part of the dipole-dipole interaction but are sensitive to interactions that take place at shorter scales. 
The contribution of short-range interaction can be modulated by tuning the applied field, making it  possible to observe a series of resonances. The angular distributions of scattered 
particles, at ultracold energies, depend sensitively on the resonance positions, and display variable degrees of anisotropy. We have also identified 
a special scattering direction in the fermionic case that is not sensitive to the short-range physics of the scattered system.

\end{document}